\documentclass[sigconf]{acmart}

\usepackage{booktabs} 

\setcopyright{rightsretained}

\begin{document}
\title{Augmenting Input Method Language Model with user Location Type Information}

\author{Di He}
\affiliation{%
  \institution{University of Illinois}
  \streetaddress{1308 W Main Street}
  \city{Urbana} 
  \state{Illinois} 
  \postcode{61801-2307}
}
\email{dihe2@illinois.edu}



\begin{abstract}
Geo-tags from micro-blog posts have been shown to be useful in many data mining applications. This work seeks to find out if the location type derived from these geo-tags can benefit input methods, which attempts to predict the next word a user will input during typing. If a correlation between different location types and a change in word distribution can be found, the location type information can be used to make the input method more accurate. This work queried micro-blog posts from Twitter API and location type of these posts from Google Place API, forming a dataset of around 500k samples. A statistical study on the word distribution found weak support for the assumption. An LSTM based prediction experiment found a 2\% edge in the accuracy from language models leveraging location type information when compared to a baseline without that information. 
\end{abstract}

%
%





\maketitle

\section{Introduction}

Micro-blog, such as Twitter and Weibo, have become a great part of modern life. Technology innovations made access to these platform possible on mobile device, for example smart phones and tablets. Due to the limitation on hardware, inputing text on mobile platforms, which usually have no dedicated key boards and limited touch screen area, is more difficult and less efficient than on other platforms. Yet, we spent a great portion of our time inputing on these devices. To assist with the text input on all them, almost all mobile device have a mechanism to assist with text input, widely called an input method. These mechanism have a built-in language model, to recommend the next word or phrase the user is most likely going to type. If the correct word or phrase is suggested, the method would be able to save the user typing time.

Early implementation of these systems have limited capability as they are built on fix and out-of-context language models. Later systems start to adapt to user inputs and adjust the language model after recording user input and preference. Still, building robust language models for micro-blog proved to be more difficult than other application~\cite{yan2015tackling, brian2016kanopy4tweets}. A collection of reasons made building good performing language models difficult for micro-blog. Just bring up a few, there exist a lot of incorrect and informal word, phrase and language habits in these blog postings. The tolerance of mistake on these platform in general is quite high for reader and writer. Not to mention these text string usually are very short~\cite{yan2015tackling}, the Micro-blog Twitter does not allow posts longer than 150 works, and larger portion of the text are name entities~\cite{brian2016kanopy4tweets}. Many attempts~\cite{yan2015tackling,tang2014learning,brian2016kanopy4tweets}, have been made to improve language model adaptation level for micro-blog text input.

Although the difficulties of building robust language model for micro-blog is challenging, we should also acknowledge, mobile platform, which is where these input methods are needed the most, also collects many other information. Among these information, users location is one that has been proved to be useful in many tasks. Geo-tagged micro-blog has been proved to be helpful in many studies. For example event detection, both on a global~\cite{atefeh2015survey,steiger2015advanced} and local~\cite{zhang2016geoburst} scale and travel recommending~\cite{xu2015topic}.

The shortage in language model and the successful examples making use of geographical information leaves us wondering if we can make use of the later to augment the former and improve its accuracy. Leveraging GPS coordinate alone is not a commonly practiced in geo-tag using. However, Geo-tags do play key rolls in user spatial and temporal patters discovery~\cite{zhang2015assembler}. This implies user wording patters do diverge according to different location or establishment shift in on a smaller scale, this shift may be from office to gym, from gym back home. We attempt to find out if this location type shifting can be leveraged to augment language model.

This report is composed of 4 remaining parts. In~\ref{sec:basic} we introduce the related concept; in~\ref{sec:method} we report the result of some statistical study and explain our techniques of making use of location type information in language modeling. After we report experimental results in~\ref{sec:res}, we conclude and discuss the work in~\ref{sec:dec}. 

\section{Background}\label{sec:basic}

In some ways, data mining is the study of knowledge finding and truth finding from noisy and unordered data sources, sometimes with overwhelming size of data. The low concentration of useful information and large quantity of data makes it a unique study of its own, separating it from traditional subjects like machine learning and database. As much as the later 2 subject is the mean and object of data mining, they can no longer incorporate the later as it has become too large and too focused on itself. Among all application of data mining, blog mining is a classic task. However, micro-blog mining, have recently attracted more attention than classic blog mining.

\subsubsection{Geo-tagged Twitter Posts}\label{sec:geo-tag}

Take Twitter as an example, study from~\cite{sloan2015tweets} found that 41.6\% of the posts on Twitter have Geo-tagged option enabled. Within all Twitter posts, only 0.85\% to 3.1\%, depending on the way posts are counted, of message have a Geo-tag embedded. 
For Twitter, Geo-tagged posts can be queried with 2 information, the user coordinate and a "place" information tagged along with the post. It is very important to separate the 2 kind of Geo-tags as they could be of totally different use to the data mining.

The first kind of Geo-tag reports the actual location of the post using the latitude and longitude collected from the hardware used to make the post. If the error from the locating hardware is ignored, this is probably the most straight forward and accurate Geo-tag. An example from~\cite{eisenstein2010latent} is given below in Fig~\ref{fig:geo_post}.

\begin{figure}[htbp]
      \centering
      \includegraphics[width=\linewidth]{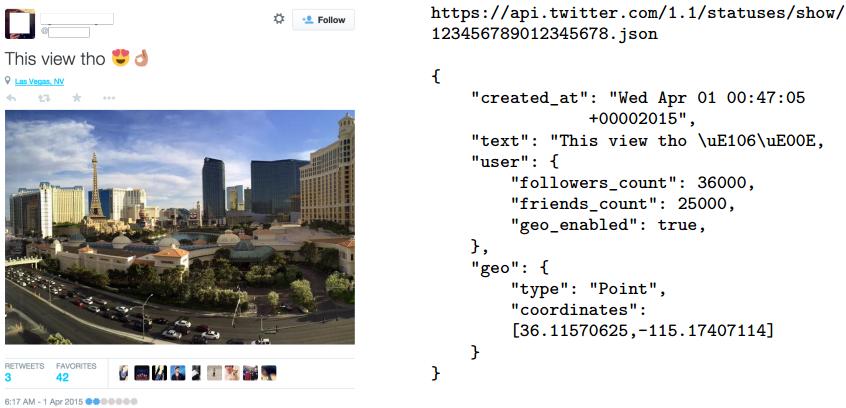}
      \caption{A Geo-tagged Twitter Post with Coordinates\label{fig:geo_post}}
\end{figure}

\begin{figure}[htbp]
      \centering
      \includegraphics[width=\linewidth]{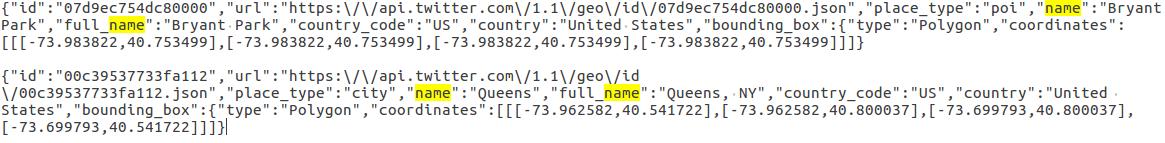}
      \caption{A Geo-tagged Twitter Post with Places\label{fig:geo_place}}
\end{figure}

The second way Twitter will match a query with location filtering is through Twitter's built-in "place" information. Each "place" in the Twitter database is a bounding box defined by latitude and longitude coordinates. These posts make up the majority of posts returned if we query with location. The problem about these returned message is that the establishment defined in the Twitter "place" database can be very large, such as a city, or small enough that the retrieved location information is as detailed as coordinates. Fig~\ref{fig:geo_place} provides an example of 2 "place" tagged along Geo-tagged Tweets.

Accurate coordinate geo-tags are essential to applications like local event detection~\cite{zhang2016geoburst}. However, the tag itself does not include any additional information, this makes it less useful in study where semantic associated with the location is required. "Place" information offers a geographic information usually in a much larger scale, studies that look at divergence that will only occur between cities and states or even nations can benefit from these information~\cite{eisenstein2010latent}. Detailed coordinates are not required in these task. For "place" with "place\_type" defined small enough, in a scale sense, not only offers similar usefulness as coordinates, it also provides additional information and semantic associated with the location.  

\subsection{Location Type}

Although Geo-tags, in different forms, are attached with many Micro-blog posts. These tags cannot be used directly to determine the function of the establishment, or the type of the location. To fully automate the process, we leverage open access APIs from Google. Google Place~\cite{PlacesAP88:online} API allows 3rd party applications to take advantage of the Google Map service and offer user location and nearby establishment searches. To provide nearby establishment search according to specific type, all establishment has multiple type tags associated with it. A list of available tags can be found at~\cite{PlaceTyp16:online}. These tags are usually specific enough to derive the function of the establishment. For example, here exists separate tags for "food\_delievery" and "food\_takeaway". However, some tags can be better organized, for example tags related to food takes multiple names including "food", "restaurant", "food\_takeaway" and "food\_deliver". On the other hand, tags like "establishment" and "point\_of\_interest" provides little information of the function of the location.

Although the location coordinates for the same establishment may vary between Google Place and Twitter. Pairing the establishment name and allowing for the Place API to search within a small radius almost always return the correct location. This allows the location type query to be fully automatic and fairly accurate.

\subsection{Language Model Adaptation}

\begin{figure}[htbp]
      \centering
      \includegraphics[width=50mm]{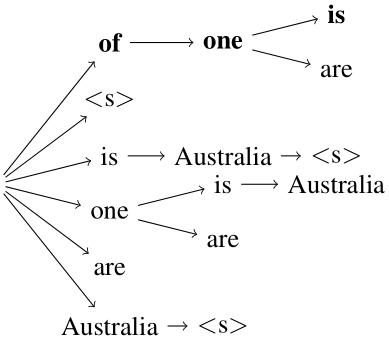}
      \caption{A Language Model\label{fig:small_lm}}
\end{figure}

Language model, also called statistical language model, encodes the word and phrase relationship within sequences of words using probability distribution models. Compared to input methods, it usually is more carefully studied by speech recognition scientists. This is because language models, which incorporates the grammar and context distribution of words are essential for Automatic Speech Recognition (ASR) systems to work~\cite{bellegarda2004statistical}. Apart from the acoustic characteristics of the speech utterance, an ASR also relays on information how a work is likely going to follow anther to make good "guesses" of the correct speech. An example from~\cite{Newlangu54:online} presents the language model in Fig~\ref{fig:small_lm}. The model only has 5 words, "of", "one", "is", "are" and "Australis", and there is a likelihood value associated with each path from left to right in the figure. Without a good language model, building large vocabulary ASR systems are almost impractical.

Language models usually records information taking the following form: Eq~\ref{eq:lang_mod}, where $Pr(w_1, ..., w_N)$ represents the probability the word sequence $w_1$, ..., $w_N$ occurs in language. That is, the probability within a word sequence of length $N$, the first word is $w_1$ and second $w_2$, ..., the last being $w_N$. 

\begin{equation}\label{eq:lang_mod}
 Pr(w_{1} , ..., w_{N})=\prod_{q=1}^{N} Pr(w_{q} | h_{q})
\end{equation}
 
 In Eq~\ref{eq:lang_mod}, the term $Pr(w_{q} | h_{q})$ represent the probability that given the words occurring before $q$, the probability of the next word following following this sequence of $q-1$ words to be $w_q$. A formal way of defining $h_q$ is Eq~\ref{eq:word_seq}.
 
\begin{equation}\label{eq:word_seq}
 h_q=w_{1}, ...,w_{q-1}
\end{equation} 
 
 Notice, if we decrease the number $q$ one-by-one from $N$ down to $2$, and we have $Pr(w_1|w_1)$ well defined, the equation is self-defined up to arbitrary length $N$. Unfortunately, when we grow the length of $h_q$, the amount of memory space we need to store all possible combinations of words increases exponentially. This is why only most recent words in $h_q$ is considered, changing Eq~\ref{eq:word_seq} into the following form:
 
\begin{equation}\label{eq:word_seqn}
 h_q=w_{q-n+1}.,w_{q-1}
\end{equation}

where in Eq~\ref{eq:word_seqn} $n$ is much smaller than $n$. In large ASR systems where memory is not of crucial concern, $n$ takes a number around $3$ to $5$. In small systems with relative large vocabulary, $n=1$, owning the language model the name of a uni-gram language model.

For an ASR systems, the decoder of the system dynamically searches the probability of all possible word sequence, jointly considers the acoustic likelihood and picks the word sequence with the highest likelihood. For an import method, the system only need to output the most likely word or the most likely set of words with the highest or one the highest $Pr(w_{q} | h_{q})$.

Despite the big table storing likelihood of all possible sequence being the most popular form language model representation. Language models based on Neural Networks have become popular, in many cases, these models have proved advantage over traditional N-gram models~\cite{chen2015recurrent}. A well trained model will still attempt to output the likelihood of a word sequence, but the underlying reasoning these likelihood numbers are generated is no longer clear or in a human readable form.

\subsubsection{Language Model Adaptation and Smoothing}

Many works have found adjusting the language model according the context, semantic or other related information can increase the ASR or input method output accuracy~\cite{bellegarda2004statistical}. In many cases, multiple language model is generated, and an heuristic or reasoning will be derived to "chose" the best one. Many different techniques have been made to select and regress between different models. We will talk about these techniques in more details. However, before that, there is one more concept we should touch on.

Language model is built or trained based on a knowledge base, this base usually is a great collection of text. However, as one could expect, some words will always occur very infrequent despite the size of the database. As a result, probability estimated for some sequence will be very unreliable or, in the worst case, be $0$ since that sequence never occurred. In order to make sense out of this situation, the original model based only on distribution will have to be "smoothed out". Some probability will have to be subtracted from frequent occurring sequences and given to rear words. According to ~\cite{yan2015tackling}, smoothing for Micro-blog is even more challenging than traditional tasks.

\section{Method}\label{sec:method}

\subsection{Dataset}

To conduct this study, Geo-tagged Twitter posts from 2 major US city, New York and San Fransisco has been collected using the Twitter API. Over $2$ million tweets have been collected, over roughly one and a half week. After that, an aggressive pruning has been carried out on the dataset. All tweets not using English, without a place tag or a place tag that is too broad, such as city, has been removed. After this, the place type for each tweet has been queried using Google Place API. The vast majority of the Twitter place can be found and matched through Google Place, but a very limited queries failed. Tweets that failed the Google Place query has been dropped from the dataset. At the end of the day, only $36$ thousand tweets was left. These Tweets contain around $507$ thousand word or special character. In actually training, additional training samples have been dropped as some tweets contain only $1$ word, which is of no use to our application.

A pre-trained dataset has been built to pre-train the baseline and a portion of the combinational network that jointly considers the previous input and the location type. This pre-train dataset is mutually independent of the training dataset mentioned above. This dataset contains only English tweets and is roughly 3 times larger than the set mentioned above.

\subsection{Formulating the Problem}

The main goal of this work is try to improve the accuracy of input methods. If we can correctly predict the next word the user intend to input most of the time, we succeeded on achieving our goal. Due to this reason, we formulated this question as a classification problem. Given the words the user has already typed in, and the information we can derive from the Geo-tags, how can we classify the word the user is going to input next. Since the support of the majority of the words are very low. Building a traditional FST based language model will encounter serious language model smoothing issue~\cite{yan2015tackling}. Since smoothing is not the focus of this study, we decide to use a neural network based language model~\cite{chen2015recurrent} instead. At the end, a classifier based on a sequential neural network is built to implement the underlying language model for the input method.

The neural network will offer us flexibility to integrate the location information in without the need of complicated model switching mechanism. It is natural to consider sequence-to-sequence~\cite{sutskever2014sequence} or attention~\cite{luong2015effective} based technique as there are showed to be promising handling text sequence. However,to limit the scope of this study, they will not be consider. The DNN based classifier will only look back a fixed number of words to conduct the prediction.

\subsubsection{Predict Class}

The training dataset mentioned above contains roughly 27 thousand different words or special characters. Special characters include Hashtags and Emojis. However, the vast majority of the these word show up only a couple of times in the entire dataset. The top $1000$ frequent words makeup over $85\%$ of the tweeting content. When the top $2000$ frequent words are selected, the covered vocabulary grow less than $5\%$. As a result, only the top $1000$ frequent word is consider in input method. All words outside of the list will be replaced with a dedicated "<unk>" label.

\subsubsection{Re-sampling Training Data}

One of the reason only the top $1000$ classes have been considered is the remaining words would have a very unbalanced support. However, this only partially solve the training data imbalance problem~\cite{japkowicz2000class}. Classic re-sampling technique has been applied to the training set to balance the dataset. 

In practice, the mean of the support of all considered words is heavily biased towards the low support words. This implies the minority of the class actually falls into the minority class category. As a result the oversampling factor of minority class has been capped at $3$ times. A minority class can be oversampled no more the $3$ times into the training set. The majority class can be under-sampled to $\mu +\sigma$, where $\mu$ is the mean of the support for all words and $\sigma$ is the standard deviation. In practice, since the mean is bias towards the minority class, the full dataset ended up shrinking slightly, to $503$ thousand samples.

\subsection{Diverge in Vocabulary}\label{sec:div}

This work assumes the language model, given the user location, will vary under different location type. However, this assumption cannot be true if the topic and context of the micro-blog post is not correlated with the user location. One can assume a person in a restaurant is more likely going to post blog on food and the restaurant. Words discussing food should have a higher likelihood been input by the user, therefore the language model should bases towards these words. However, it is not guaranteed that this significantly shifts the likelihood of words. It could be the case that most user in the restaurant do not comment on their food. Even though first-visitors to a nice dining location may be interested in sharing their thoughts on the location, expressing the same interests from returning customers do not sound very attractive. Another reason this assumption might fall apart is the shift in vocabulary distribution in specific location or establishments might not be distinctive enough. One can image tourists might favor a vocabulary different from others when they are visiting a museum. However, it is hard to image how choices of words will diverge significantly when a user is in a subway station or bus stop.

All these doubt calls for a statistical study to back up the hypothesis that assumes people's vocabulary is correlated to their location type. Due to this reason, we conducted a Chi-square~\cite{moore1976chi} test on the word distribution of each different location types. The Chi-square score of the most significant and most in-significant 10 locations are listed below in Tab~\ref{tab:chi}. In the study, a word has to collect a minimum 5 support within a location type to be considered. Locations without any words satisfying the minimum support will be dropped from the study.

\begin{table}[htbp]
\centering
\caption{Chi-square score for Different Location Type}
\label{tab:chi}
\begin{tabular}{ll|ll}
\multicolumn{2}{l|}{Top 10 Places}   & \multicolumn{2}{l}{Least 10 Places}  \\ \hline
\multicolumn{1}{l|}{Place} & Chi-squ & \multicolumn{1}{l|}{Place} & Chi-squ \\ \hline
embassy                    & 213.219 & accounting                 & 0.135   \\
department\_store          & 8.721   & home\_goods\_store         & 0.125   \\
premise                    & 5.177   & atm                        & 0.117   \\
shoe\_store                & 4.650   & finance                    & 0.110   \\
jewelry\_store             & 4.148   & store                      & 0.107   \\
church                     & 3.741   & bus\_station               & 0.102   \\
train\_station             & 3.541   & bank                       & 0.097   \\
hospital                   & 2.694   & aquarium                   & 0.089   \\
stadium                    & 2.623   & rv\_park                   & 0.072   \\
place\_of\_worship         & 2.192   & laundry                    & 0.061  
\end{tabular}
\end{table}

The average Chi-square score for 62 considered locations is $16.23$ and 16 of the places have a score above $1$. We can see, the Chi-square does imply posts from some locations have observable bias in choices of words. However, the shift in distribution for most locations are not very significant. In Tab~\ref{tab:sig_word}, we listed some words with significant shift in distribution for a couple of locations. It can be seen that word with high, above $1$, medium, between $1$ and $0.5$, Chi-square score have words that have semantic meanings correlated with the location. But locations with low score seems to lack meaningful word in general.

\begin{table}[htbp]
\centering
\caption{Significant Words from some Locations}
\label{tab:sig_word}
\begin{tabular}{lll}
premise   & food   & rv\_park \\ \hline
trump     & dinner & if       \\
tower     & lunch  & think    \\
protest   & svu    & )        \\
midtown   & foodie & would    \\
manhattan & brunch & one      \\
nyc       & menu   & he  	  \\
)         & cream   & as       \\
(         & bayarea & your  
\end{tabular}
\end{table}

\subsection{Network Structure}

\begin{figure}[htbp]
      \centering
      \includegraphics[width=60mm]{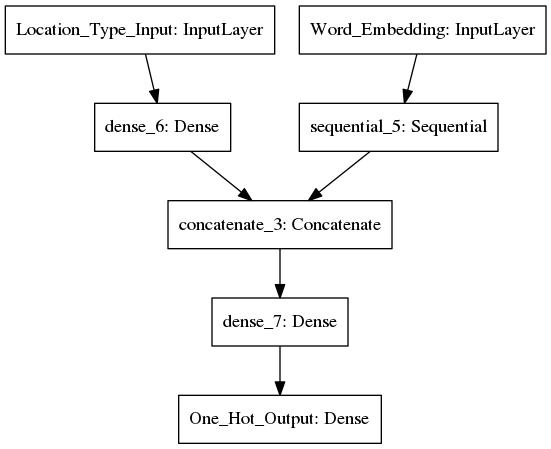}
      \caption{Overall Structure of the Neural Network\label{fig:full_nn}}
\end{figure}

A network structure has been proposed in Fig~\ref{fig:full_nn}. Although we chose to drop a list of words from the classifier output as failing to predict them will not offset the performance dramatically, we do not want to miss low-frequency words from the input. Yet, due to the sparse nature of text~\cite{mikolov2013efficient}, it is hard to effectively represent words using low dimension vector, until the recently introduced word-to-vector embedding technique~\cite{mikolov2013efficient}. This is why for the network branch that handles the previous words, an embedding layer is first used to embed the words into a low dimension vector. In practice, training a good word embedding layer can be time and resource consuming, and these layer are context sensitive. Meaning a embedding trained on Twitter dataset might not be very ideal for newspaper, vise-versa. In order to obtain good word embedding, we borrowed the GloVe embedding set for Twitter~\cite{pennington2014glove}. This already trained embedding is one of the most highly cited word embedding word, the Twitter embedding contains a vocabulary of $27$ billion words or special characters collected over a couple of month. The same vocabulary is embedded into $25$, $50$, $100$ and $200$ dimensional vectors. Introducing an embedding layer benefits the network by allowing it effectively encode a large vocabulary. However, it also introduces a level of in-transparency as we have to consider the proximity of each word in the embedding space. This proximity might not be correlated with the location type very well. We will discuss this issue in more details in~\ref{sec:emb}.

\begin{figure}[htbp]
      \centering
      \includegraphics[width=60mm]{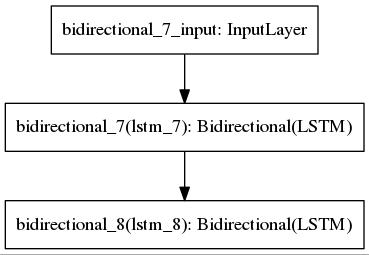}
      \caption{Sequential Structure of the Neural Network\label{fig:seq_nn}}
\end{figure}

The core network portion that handles previous word input is build up by a 2-layer bidirectional Long-short Term Memory (LSTM) RNN~\ref{fig:seq_nn}.

LSTM has been shown to handle sequential input very well in speech recognition~\cite{graves2005framewise}. It is believed to be able to remember information from further past when compared to a classic RNNs. Later work leveraged the advantage of LSTM to handle sequence input in all form, language model is no doubt one of the application that benefited~\cite{chen2015recurrent}. As LSTM first shown its promise in speech recognition, bi-directional LSTM networks have shown advantage over traditional 1-directional versions~\cite{graves2005framewise}. In a bi-directional network, one layer has been doubled in size, and half the network handles the input in its original order, the other half of the network handles the same input sequence in reverse order. After some empirical study, it has been found that 2 recurrent layers, with decent node count reaches a high accuracy.  

Location types have been encoded into multi-hot vectors. These are input similar to 1-hot vector, however, multiple dimension can have non-zero value since a place will have multiple definition in the query result from Google Place. A fully connected layer, namely dense\_6 in Fig~\ref{fig:full_nn}, has been added between the place input and the concatenate layer to project the place input to a smaller dimension. A fully connected layer, dense\_7 in Fig~\ref{fig:full_nn}, is added to provide the network more flexibility to handle the frequency shift from each location type. 

\subsubsection{Is the Location-based Frequency Bias Significant After the Embedding?}\label{sec:emb}

As embedding layer has been applied to the network, we are concerned with one possibility. Has the distribution shift according to location type propagated into the embedded word space. According to the way wards are embedded in the word-to-vector setup, words that co-occur in close proximity should be embedded close to each other in the embedded space~\cite{mikolov2013efficient}. In theory this implies words that share the same location type should be embedded close to each other as they most likely will share same appearances in posts. However, this assumption ignores the case that a word may be popular for multiple diverge locations. For example the same word "up" turned out to be significant at bank and department\_store from the analysis in~\ref{sec:div}. It is hard to image other words frequent from these 2 location embedded together.

To study how closely words are embedded together according to their location type. We randomly sampled $200$ words from each location type with decent support, meaning have to have a word count support at least $5$ times the size of the sample count, and calculated their average standard deviation among all the dimension of its embedding vector. The chance a word is drawn is positively correlated to the support of that word. The assumption is if words from a location type is clustered together, the average standard deviation from the sampling should be low compared to randomly sampling words without considering their location type. If the words from the same location type indeed clustered together in the embedding space, the network can more easily promote these words when location information is presented. If it is the other way around, the network will have a hard time making use of the location type information as the embedding vector for each location is scattered in the embedding space. The embedding layer will effectively randomize the input for words belonging to the same location type. Tab~\ref{tab:div_emb} presented the average standard deviation for some location types. The $100$ dimensional vector is used and the locations with the smallest and larges standard deviation is presented. The standard deviation from random sampling is $0.444$.

\begin{table}[htbp]
\centering
\caption{Average Standard Deviation of Word Embedding Vector from Different Location Types}
\label{tab:div_emb}
\begin{tabular}{ll|ll}
\multicolumn{2}{l|}{Lowest STD Place} & \multicolumn{2}{l}{Highest STD Place} \\ \hline
\multicolumn{1}{l|}{Place}   & STD    & \multicolumn{1}{l|}{Place}   & STD    \\ \hline
electronics\_store           & 0.315  & art\_gallery                 & 0.424  \\
book\_store                  & 0.332  & light\_rail\_station         & 0.428  \\
finance                      & 0.346  & place\_of\_worship           & 0.429  \\
meal\_takeaway               & 0.359  & beauty\_salon                & 0.430  \\
rv\_park                     & 0.360  & establishment                & 0.430  \\
storage                      & 0.368  & hair\_care                   & 0.431  \\
hardware\_store              & 0.374  & museum                       & 0.433  \\
bus\_station                 & 0.374  & point\_of\_interest          & 0.438  \\
movie\_theater               & 0.374  & church                       & 0.439  \\
liquor\_store                & 0.380  & stadium                      & 0.441 
\end{tabular}
\end{table}

As we can see from the results presented in Tab~\ref{tab:div_emb}, the frequent words do not seem to cluster together in the embedded space. The standard deviation of many locations, including locations that appeared to be out-standing the Chi-square test has relatively high standard deviation. locations such as rv\_park, which scored low on the Chi-square test were, in-contrast, clustered relatively close compared to other location types. The standard deviation of many location types are very close to the standard deviation of random sampling. This is a bad indication. It may not imply the location type information will not benefit the embedded words, but is does mean the network will have to work relatively hard to associate different words to the same location type.   

\section{Experimental Results}

\subsection{Experiment Setup}

The statistical study returned mixed results. In order to fully evaluate the effectiveness of location type information, we conducted a prediction experiment with neural networks leveraging location type information, comparing it against a location information free baseline. The models of interests attempts to correctly predict the next word using location type and 4 previous words. The baseline setup does not take location type information into consideration, the network is a classic 4-gram language model. 

The {\bf baseline model} differs to the structure presented in Fig~\ref{fig:full_nn} as it does not have the place branch stretching from place\_input to concatenate\_3. The words or special character has been embedded in to $100$ dimensional vector using the embedding from GloVe~\cite{pennington2014glove}. The 2 recurrent layers are each made of up a bi-directional LSTM layer. Each direction of the LSTM contains $256$ memory cells. The recurrent portion of the network feeds into a fully connected layer with $256$ nodes. The Activation function of the node is hyperbolic-tangent (Tanh). The output of the fully connected network feeds into an output layer with $1002$ output classes. The output layer has a Softmax activation function. Among the $1002$ output classes, $1000$ nodes represent distinct words and $1$ node represent words outside of the $1000$ most frequent word set. The last node is reserved for padding. 

The baseline is compared against 3 different setup. {\bf Setup 1} skips dense\_6 layer and feeds the multi-hot place input vector directly into the concatenating layer. All $94$ location type that occurred in the dataset have a dedicated dimension in the place\_input layer. In {\bf setup 2}, dense\_6 has been removed just as setup 1 but only $62$ frequently occurring location type have dedicated dimension in place\_input. This makes place\_input a $62$ dimensional layer. If a location type not belonging to this frequent location type list is present, it will be ignored. In {\bf setup 3}, place\_input shares the same setup as {\bf setup 2}, but dense\_6 is present between input\_place and concatenate\_3. In this setup, dense\_6 is a fully connective layer with $16$ nodes.

The network without the place branch has be initialized with weights from a network with identical structure as the baseline, the later is trained on the pre-train dataset mentioned in~\ref{sec:method}. The remaining portion of the network for setups with place input has bee initialized to have the same mean and variance assuming the weights follow normal distribution. Using pre-train weights have been found to speedup the network convergence rate.

The Networks are trained using Tensorflow~\cite{abadi2016tensorflow} through the Keras~\cite{chollet2015keras} wrapper. Different network setup has been trained through the 500k dataset with 10\% of the data left out for test evaluation.

\subsection{Results}\label{sec:res}

The accuracy of the prediction after $20$ epoch has been reported in Tab~\ref{tab:nn_acc}. {\bf Top 1} reports the accuracy where only the most likely word from the output vector is compared against the target. The {\bf Top 5} result considers a sample correctly classified if the target class is within the top $5$ most likely output of the network. Consider the task is to classify a correct word out of $1000$ classes, the baseline is providing decent accuracy. It can be seen that the network taking advantage of location type information does have minor accuracy edge over the baseline, which does not consider location type. However, the difference, in both the {\bf Top 1} and {\bf Top 5} case, is only about 2\%. Within the place type setups, {\bf Setup 3} with dense\_6 between place\_input and the concatenation layer seems to under-perform the rest, lagging by 1\% in the final accuracy. 

\begin{table}[htbp]
\centering
\caption{Prediction Accuracy for Different Input and Network Setup}
\label{tab:nn_acc}
\begin{tabular}{|l|ll}
\hline
Acc (\%) & \multicolumn{1}{l|}{Top 1} & \multicolumn{1}{l|}{Top 5} \\ \hline
baseline & 63.26                      & 76.07                      \\ \cline{1-1}
setup 1  & 65.64                      & 78.24                      \\ \cline{1-1}
setup 2  & 65.83                      & 78.13                      \\ \cline{1-1}
setup 3  & 64.15                      & 79.82                      \\ \cline{1-1}
\end{tabular}
\end{table}

\begin{figure}[htbp]
      \centering
      \includegraphics[width=90mm]{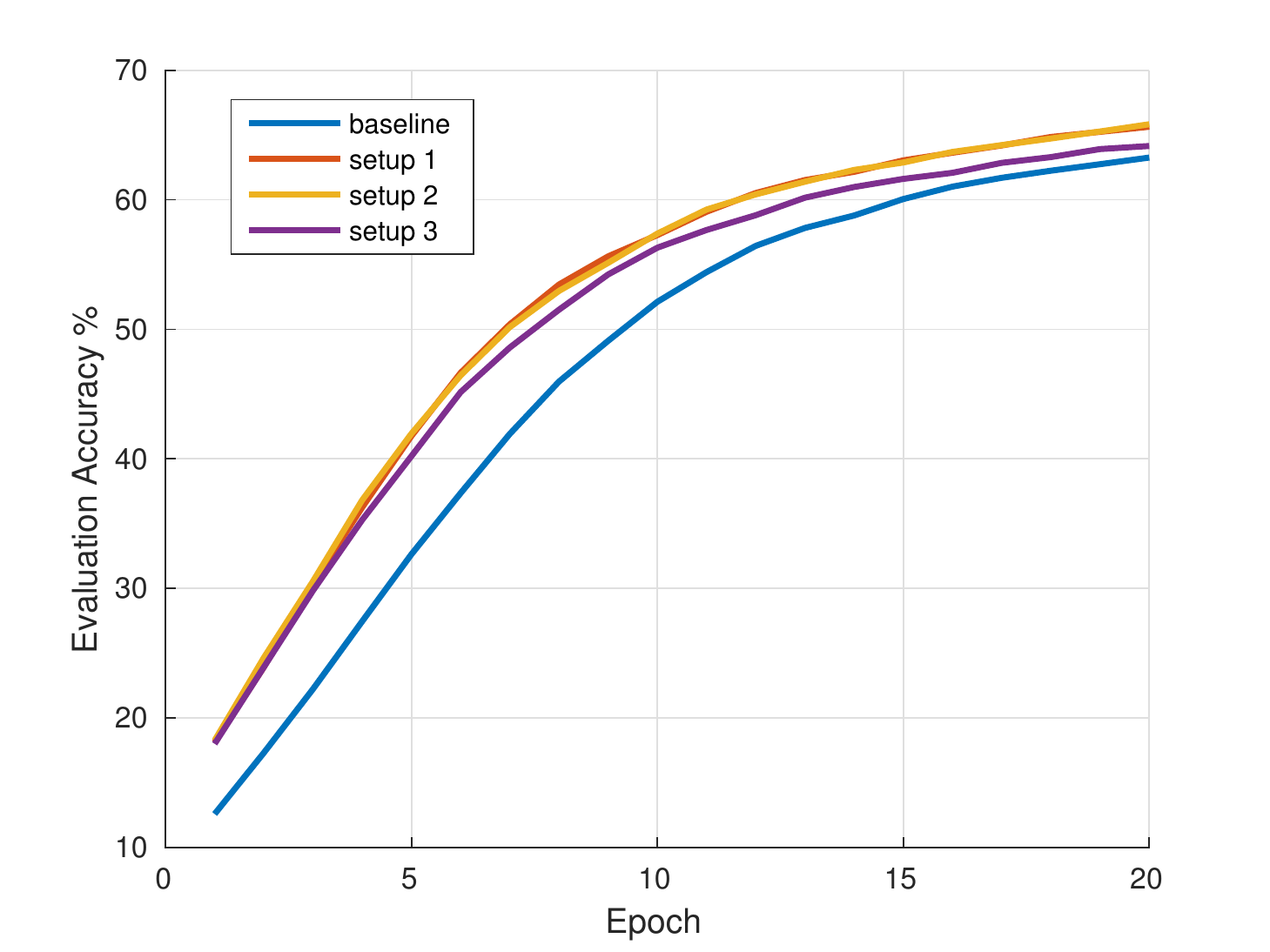}
      \caption{Convergence Rate of Different Input and Network Setup\label{fig:conv_rate}}
\end{figure}

The validation accuracy after each epoch has been reported in Fig~\ref{fig:conv_rate}. It can be seen that the setup with place input leads in accuracy throughout the training. The accuracy edge is consistent for all place setup starting at the first epoch. The accuracy advantage for place type setups increases after $5$ to $10$ epoch, however it reduces as the network starts to converge. In epochs near the end, the accuracy advantage stables around 2\% for the $2$ place setup without dense\_6. The accuracy of {\bf Setup 3} cannot catchup with the other 2 setups and only leads the baseline about 1\% at the end.

\section{Conclusions and Discussion}\label{sec:dec}

It can be seen from the results presented in~\ref{sec:res} that using location type information is capable of improving the language model accuracy when compared to a baseline, which does not consider location type. However, even through the accuracy edge is consistent, the accuracy difference is very minor. It is difficult to conclude the difference is significant to support our hypothesis that location information can effectively augment input methods. Also the difference is also too small to justify introducing the extra complexity of collecting the place type information. On the other hand, the extra information provided from location type does seems to be contributing the language model accuracy. Therefore a different technique of using this information or new network or embedding structure may return better results.

From the statistic analysis in~\ref{sec:method} we can see that the location type is weakly correlated to the word distribution change. This is understandable considering the case many people might not be posting blogs related to their location at all. This calls for the need of an extra layer of mechanism to verify the topic of the discussion and see if it is indeed correlated with the location type. Maybe conducting topic detection in combination with location type might improve the accuracy. On the other hand study from~\ref{sec:method} also indicate that the GloVe embedding may not fit the location type very ideal. Training a new embedding layer or restricting the embedding with location types might be a way to address this issue.






\begin{thebibliography}{00}


\ifx \showCODEN    \undefined \def \showCODEN     #1{\unskip}     \fi
\ifx \showDOI      \undefined \def \showDOI       #1{#1}\fi
\ifx \showISBNx    \undefined \def \showISBNx     #1{\unskip}     \fi
\ifx \showISBNxiii \undefined \def \showISBNxiii  #1{\unskip}     \fi
\ifx \showISSN     \undefined \def \showISSN      #1{\unskip}     \fi
\ifx \showLCCN     \undefined \def \showLCCN      #1{\unskip}     \fi
\ifx \shownote     \undefined \def \shownote      #1{#1}          \fi
\ifx \showarticletitle \undefined \def \showarticletitle #1{#1}   \fi
\ifx \showURL      \undefined \def \showURL       {\relax}        \fi
\providecommand\bibfield[2]{#2}
\providecommand\bibinfo[2]{#2}
\providecommand\natexlab[1]{#1}
\providecommand\showeprint[2][]{arXiv:#2}

\bibitem[\protect\citeauthoryear{Abadi, Agarwal, Barham, Brevdo, Chen, Citro,
  Corrado, Davis, Dean, Devin, et~al\mbox{.}}{Abadi et~al\mbox{.}}{2016}]%
        {abadi2016tensorflow}
\bibfield{author}{\bibinfo{person}{Mart{\'\i}n Abadi}, \bibinfo{person}{Ashish
  Agarwal}, \bibinfo{person}{Paul Barham}, \bibinfo{person}{Eugene Brevdo},
  \bibinfo{person}{Zhifeng Chen}, \bibinfo{person}{Craig Citro},
  \bibinfo{person}{Greg~S Corrado}, \bibinfo{person}{Andy Davis},
  \bibinfo{person}{Jeffrey Dean}, \bibinfo{person}{Matthieu Devin}, {and}
  \bibinfo{person}{others}.} \bibinfo{year}{2016}\natexlab{}.
\newblock \showarticletitle{Tensorflow: Large-scale machine learning on
  heterogeneous distributed systems}.
\newblock \bibinfo{journal}{{\em arXiv preprint arXiv:1603.04467\/}}
  (\bibinfo{year}{2016}).
\newblock


\bibitem[\protect\citeauthoryear{Atefeh and Khreich}{Atefeh and
  Khreich}{2015}]%
        {atefeh2015survey}
\bibfield{author}{\bibinfo{person}{Farzindar Atefeh} {and}
  \bibinfo{person}{Wael Khreich}.} \bibinfo{year}{2015}\natexlab{}.
\newblock \showarticletitle{A survey of techniques for event detection in
  twitter}.
\newblock \bibinfo{journal}{{\em Computational Intelligence\/}}
  \bibinfo{volume}{31}, \bibinfo{number}{1} (\bibinfo{year}{2015}),
  \bibinfo{pages}{132--164}.
\newblock


\bibitem[\protect\citeauthoryear{Bellegarda}{Bellegarda}{2004}]%
        {bellegarda2004statistical}
\bibfield{author}{\bibinfo{person}{Jerome~R Bellegarda}.}
  \bibinfo{year}{2004}\natexlab{}.
\newblock \showarticletitle{Statistical language model adaptation: review and
  perspectives}.
\newblock \bibinfo{journal}{{\em Speech communication\/}} \bibinfo{volume}{42},
  \bibinfo{number}{1} (\bibinfo{year}{2004}), \bibinfo{pages}{93--108}.
\newblock


\bibitem[\protect\citeauthoryear{Brian and Hayes}{Brian and Hayes}{2016}]%
        {brian2016kanopy4tweets}
\bibfield{author}{\bibinfo{person}{Pablo Torres-Tram{\'o}n Hugo~Hromic Brian}
  {and} \bibinfo{person}{Walsh Bahareh R Heravi~Conor Hayes}.}
  \bibinfo{year}{2016}\natexlab{}.
\newblock \showarticletitle{Kanopy4Tweets: Entity Extraction and Linking for
  Twitter}.
\newblock  (\bibinfo{year}{2016}).
\newblock


\bibitem[\protect\citeauthoryear{Chen, Tan, Liu, Lanchantin, Wan, Gales, and
  Woodland}{Chen et~al\mbox{.}}{2015}]%
        {chen2015recurrent}
\bibfield{author}{\bibinfo{person}{Xie Chen}, \bibinfo{person}{Tian Tan},
  \bibinfo{person}{Xunying Liu}, \bibinfo{person}{Pierre Lanchantin},
  \bibinfo{person}{Moquan Wan}, \bibinfo{person}{Mark~JF Gales}, {and}
  \bibinfo{person}{Philip~C Woodland}.} \bibinfo{year}{2015}\natexlab{}.
\newblock \showarticletitle{Recurrent neural network language model adaptation
  for multi-genre broadcast speech recognition.}. In \bibinfo{booktitle}{{\em
  INTERSPEECH}}, Vol.~\bibinfo{volume}{15}. \bibinfo{pages}{3511--3515}.
\newblock


\bibitem[\protect\citeauthoryear{Chollet}{Chollet}{2015}]%
        {chollet2015keras}
\bibfield{author}{\bibinfo{person}{Fran{\c{c}}ois Chollet}.}
  \bibinfo{year}{2015}\natexlab{}.
\newblock \bibinfo{title}{Keras}.
\newblock   (\bibinfo{year}{2015}).
\newblock


\bibitem[\protect\citeauthoryear{CMU}{CMU}{2017}]%
        {Newlangu54:online}
\bibfield{author}{\bibinfo{person}{CMU}.} \bibinfo{year}{2017}\natexlab{}.
\newblock \bibinfo{title}{New language model binary format « CMU Sphinx}.
\newblock
  \bibinfo{howpublished}{\url{http://cmusphinx.sourceforge.net/2015/07/new-language-model-binary-format/}}.
    (\bibinfo{date}{May} \bibinfo{year}{2017}).
\newblock
\newblock
\shownote{(Accessed on 04/08/2017).}


\bibitem[\protect\citeauthoryear{Eisenstein, O'Connor, Smith, and
  Xing}{Eisenstein et~al\mbox{.}}{2010}]%
        {eisenstein2010latent}
\bibfield{author}{\bibinfo{person}{Jacob Eisenstein}, \bibinfo{person}{Brendan
  O'Connor}, \bibinfo{person}{Noah~A Smith}, {and} \bibinfo{person}{Eric~P
  Xing}.} \bibinfo{year}{2010}\natexlab{}.
\newblock \showarticletitle{A latent variable model for geographic lexical
  variation}. In \bibinfo{booktitle}{{\em Proceedings of the 2010 Conference on
  Empirical Methods in Natural Language Processing}}. Association for
  Computational Linguistics, \bibinfo{pages}{1277--1287}.
\newblock


\bibitem[\protect\citeauthoryear{Google}{Google}{2017a}]%
        {PlaceTyp16:online}
\bibfield{author}{\bibinfo{person}{Google}.} \bibinfo{year}{2017}\natexlab{a}.
\newblock \bibinfo{title}{Place Types | Google Places API | Google Developers}.
\newblock
  \bibinfo{howpublished}{\url{https://developers.google.com/places/supported_types}}.
    (\bibinfo{date}{May} \bibinfo{year}{2017}).
\newblock
\newblock
\shownote{(Accessed on 05/06/2017).}


\bibitem[\protect\citeauthoryear{Google}{Google}{2017b}]%
        {PlacesAP88:online}
\bibfield{author}{\bibinfo{person}{Google}.} \bibinfo{year}{2017}\natexlab{b}.
\newblock \bibinfo{title}{Places APIs and Related Products | Google Places API
  | Google Developers}.
\newblock
  \bibinfo{howpublished}{\url{https://developers.google.com/places/documentation/}}.
    (\bibinfo{date}{May} \bibinfo{year}{2017}).
\newblock
\newblock
\shownote{(Accessed on 05/06/2017).}


\bibitem[\protect\citeauthoryear{Graves and Schmidhuber}{Graves and
  Schmidhuber}{2005}]%
        {graves2005framewise}
\bibfield{author}{\bibinfo{person}{Alex Graves} {and}
  \bibinfo{person}{J{\"u}rgen Schmidhuber}.} \bibinfo{year}{2005}\natexlab{}.
\newblock \showarticletitle{Framewise phoneme classification with bidirectional
  LSTM and other neural network architectures}.
\newblock \bibinfo{journal}{{\em Neural Networks\/}} \bibinfo{volume}{18},
  \bibinfo{number}{5} (\bibinfo{year}{2005}), \bibinfo{pages}{602--610}.
\newblock


\bibitem[\protect\citeauthoryear{Japkowicz}{Japkowicz}{2000}]%
        {japkowicz2000class}
\bibfield{author}{\bibinfo{person}{Nathalie Japkowicz}.}
  \bibinfo{year}{2000}\natexlab{}.
\newblock \showarticletitle{The class imbalance problem: Significance and
  strategies}. In \bibinfo{booktitle}{{\em Proc. of the Int’l Conf. on
  Artificial Intelligence}}.
\newblock


\bibitem[\protect\citeauthoryear{Luong, Pham, and Manning}{Luong
  et~al\mbox{.}}{2015}]%
        {luong2015effective}
\bibfield{author}{\bibinfo{person}{Minh-Thang Luong}, \bibinfo{person}{Hieu
  Pham}, {and} \bibinfo{person}{Christopher~D Manning}.}
  \bibinfo{year}{2015}\natexlab{}.
\newblock \showarticletitle{Effective approaches to attention-based neural
  machine translation}.
\newblock \bibinfo{journal}{{\em arXiv preprint arXiv:1508.04025\/}}
  (\bibinfo{year}{2015}).
\newblock


\bibitem[\protect\citeauthoryear{Mikolov, Chen, Corrado, and Dean}{Mikolov
  et~al\mbox{.}}{2013}]%
        {mikolov2013efficient}
\bibfield{author}{\bibinfo{person}{Tomas Mikolov}, \bibinfo{person}{Kai Chen},
  \bibinfo{person}{Greg Corrado}, {and} \bibinfo{person}{Jeffrey Dean}.}
  \bibinfo{year}{2013}\natexlab{}.
\newblock \showarticletitle{Efficient estimation of word representations in
  vector space}.
\newblock \bibinfo{journal}{{\em arXiv preprint arXiv:1301.3781\/}}
  (\bibinfo{year}{2013}).
\newblock


\bibitem[\protect\citeauthoryear{Moore}{Moore}{1976}]%
        {moore1976chi}
\bibfield{author}{\bibinfo{person}{David~S Moore}.}
  \bibinfo{year}{1976}\natexlab{}.
\newblock \bibinfo{booktitle}{{\em Chi-Square Tests.}}
\newblock \bibinfo{type}{{T}echnical {R}eport}. \bibinfo{institution}{DTIC
  Document}.
\newblock


\bibitem[\protect\citeauthoryear{Pennington, Socher, and Manning}{Pennington
  et~al\mbox{.}}{2014}]%
        {pennington2014glove}
\bibfield{author}{\bibinfo{person}{Jeffrey Pennington},
  \bibinfo{person}{Richard Socher}, {and} \bibinfo{person}{Christopher~D
  Manning}.} \bibinfo{year}{2014}\natexlab{}.
\newblock \showarticletitle{Glove: Global Vectors for Word Representation.}. In
  \bibinfo{booktitle}{{\em EMNLP}}, Vol.~\bibinfo{volume}{14}.
  \bibinfo{pages}{1532--1543}.
\newblock


\bibitem[\protect\citeauthoryear{Sloan and Morgan}{Sloan and Morgan}{2015}]%
        {sloan2015tweets}
\bibfield{author}{\bibinfo{person}{Luke Sloan} {and} \bibinfo{person}{Jeffrey
  Morgan}.} \bibinfo{year}{2015}\natexlab{}.
\newblock \showarticletitle{Who tweets with their location? understanding the
  relationship between demographic characteristics and the use of geoservices
  and geotagging on twitter}.
\newblock \bibinfo{journal}{{\em PloS one\/}} \bibinfo{volume}{10},
  \bibinfo{number}{11} (\bibinfo{year}{2015}), \bibinfo{pages}{e0142209}.
\newblock


\bibitem[\protect\citeauthoryear{Steiger, Albuquerque, and Zipf}{Steiger
  et~al\mbox{.}}{2015}]%
        {steiger2015advanced}
\bibfield{author}{\bibinfo{person}{Enrico Steiger},
  \bibinfo{person}{Jo{\~a}o~Porto Albuquerque}, {and}
  \bibinfo{person}{Alexander Zipf}.} \bibinfo{year}{2015}\natexlab{}.
\newblock \showarticletitle{An advanced systematic literature review on
  spatiotemporal analyses of Twitter data}.
\newblock \bibinfo{journal}{{\em Transactions in GIS\/}} \bibinfo{volume}{19},
  \bibinfo{number}{6} (\bibinfo{year}{2015}), \bibinfo{pages}{809--834}.
\newblock


\bibitem[\protect\citeauthoryear{Sutskever, Vinyals, and Le}{Sutskever
  et~al\mbox{.}}{2014}]%
        {sutskever2014sequence}
\bibfield{author}{\bibinfo{person}{Ilya Sutskever}, \bibinfo{person}{Oriol
  Vinyals}, {and} \bibinfo{person}{Quoc~V Le}.}
  \bibinfo{year}{2014}\natexlab{}.
\newblock \showarticletitle{Sequence to sequence learning with neural
  networks}. In \bibinfo{booktitle}{{\em Advances in neural information
  processing systems}}. \bibinfo{pages}{3104--3112}.
\newblock


\bibitem[\protect\citeauthoryear{Tang, Wei, Yang, Zhou, Liu, and Qin}{Tang
  et~al\mbox{.}}{2014}]%
        {tang2014learning}
\bibfield{author}{\bibinfo{person}{Duyu Tang}, \bibinfo{person}{Furu Wei},
  \bibinfo{person}{Nan Yang}, \bibinfo{person}{Ming Zhou},
  \bibinfo{person}{Ting Liu}, {and} \bibinfo{person}{Bing Qin}.}
  \bibinfo{year}{2014}\natexlab{}.
\newblock \showarticletitle{Learning Sentiment-Specific Word Embedding for
  Twitter Sentiment Classification.}. In \bibinfo{booktitle}{{\em ACL (1)}}.
  \bibinfo{pages}{1555--1565}.
\newblock


\bibitem[\protect\citeauthoryear{Xu, Chen, and Chen}{Xu et~al\mbox{.}}{2015}]%
        {xu2015topic}
\bibfield{author}{\bibinfo{person}{Zhenxing Xu}, \bibinfo{person}{Ling Chen},
  {and} \bibinfo{person}{Gencai Chen}.} \bibinfo{year}{2015}\natexlab{}.
\newblock \showarticletitle{Topic based context-aware travel recommendation
  method exploiting geotagged photos}.
\newblock \bibinfo{journal}{{\em Neurocomputing\/}}  \bibinfo{volume}{155}
  (\bibinfo{year}{2015}), \bibinfo{pages}{99--107}.
\newblock


\bibitem[\protect\citeauthoryear{Yan, Li, Liu, and Hu}{Yan
  et~al\mbox{.}}{2015}]%
        {yan2015tackling}
\bibfield{author}{\bibinfo{person}{Rui Yan}, \bibinfo{person}{Xiang Li},
  \bibinfo{person}{Mengwen Liu}, {and} \bibinfo{person}{Xiaohua Hu}.}
  \bibinfo{year}{2015}\natexlab{}.
\newblock \showarticletitle{Tackling Sparsity, the Achilles Heel of Social
  Networks: Language Model Smoothing via Social Regularization.}. In
  \bibinfo{booktitle}{{\em ACL (2)}}. \bibinfo{pages}{623--629}.
\newblock


\bibitem[\protect\citeauthoryear{Zhang, Zheng, Ma, and Han}{Zhang
  et~al\mbox{.}}{2015}]%
        {zhang2015assembler}
\bibfield{author}{\bibinfo{person}{Chao Zhang}, \bibinfo{person}{Yu Zheng},
  \bibinfo{person}{Xiuli Ma}, {and} \bibinfo{person}{Jiawei Han}.}
  \bibinfo{year}{2015}\natexlab{}.
\newblock \showarticletitle{Assembler: efficient discovery of spatial
  co-evolving patterns in massive geo-sensory data}. In
  \bibinfo{booktitle}{{\em Proceedings of the 21th ACM SIGKDD International
  Conference on Knowledge Discovery and Data Mining}}. ACM,
  \bibinfo{pages}{1415--1424}.
\newblock


\bibitem[\protect\citeauthoryear{Zhang, Zhou, Yuan, Zhuang, Zheng, Kaplan,
  Wang, and Han}{Zhang et~al\mbox{.}}{2016}]%
        {zhang2016geoburst}
\bibfield{author}{\bibinfo{person}{Chao Zhang}, \bibinfo{person}{Guangyu Zhou},
  \bibinfo{person}{Quan Yuan}, \bibinfo{person}{Honglei Zhuang},
  \bibinfo{person}{Yu Zheng}, \bibinfo{person}{Lance Kaplan},
  \bibinfo{person}{Shaowen Wang}, {and} \bibinfo{person}{Jiawei Han}.}
  \bibinfo{year}{2016}\natexlab{}.
\newblock \showarticletitle{GeoBurst: Real-Time Local Event Detection in
  Geo-Tagged Tweet Streams}. In \bibinfo{booktitle}{{\em Proceedings of the
  39th International ACM SIGIR conference on Research and Development in
  Information Retrieval}}. ACM, \bibinfo{pages}{513--522}.
\newblock


\end{thebibliography}

\end{document}